# Nachhaltige Strategien gegen die COVID-19-Pandemie in Deutschland im Winter 2021/2022

## Abstract

In this position paper, a large group of interdisciplinary experts outlines response strategies against the spread of SARS-CoV-2 in the winter of 2021/2022 in Germany. We review the current state of the COVID-19 pandemic, from incidence and vaccination efficacy to hospital capacity. Building on this situation assessment, we illustrate various possible scenarios for the winter, and detail the mechanisms and effectiveness of the non-pharmaceutical interventions, vaccination, and booster. With this assessment, we want to provide orientation for decision makers about the progress and mitigation of COVID-19.

## Zusammenfassung

Seit Juli 2021 ist auch in Deutschland die Delta-Variante des SARS-Coronavirus-2 (SARS-CoV-2) dominant. Sie ist deutlich ansteckender als die bisher bekannten Varianten. Das bedeutet ein hohes Risiko für Ungeimpfte, sich anzustecken. Dazu kommt, dass der Impfschutz ohne Dritt-Impfung ("Booster") mit der Zeit kontinuierlich nachlässt. Nach fünf Monaten sinkt der Schutz gegen Ansteckung von einem Faktor von etwa 10 auf einen Faktor 2-3. Dadurch tragen aktuell nicht nur die ungeimpften Personen zur Ausbreitung des Virus bei, sondern auch jener Teil der Bevölkerung, bei dem die Impfung schon länger zurückliegt. Zusammen mit der Saisonalität schlägt sich das in einem rapiden Anstieg der Inzidenzen nieder.

Die aktuelle Situation ist dadurch sehr kritisch. Die exponentiell steigenden Inzidenzen sind direkt mit der Belegung auf den Intensivstationen gekoppelt. Um einer dauerhaften Überlastung der Intensivstationen entgegenzuwirken, ist insbesondere das Impfen und die Verbesserung des Impfschutzes durch eine Dritt-Impfung nachhaltig wirksam. Impfungen haben zwei Schutzwirkungen: Sie schützen die geimpfte Person vor schwerer Erkrankung und ihre Umgebung vor einer Übertragung. Mit jeder Verbesserung des Impfschutzes wird also sowohl die Ausbreitung als auch die Wahrscheinlichkeit eines schweren Verlaufes reduziert. Das Boostern und das Schließen der Impflücken ist deswegen extrem hilfreich. Aktuell stehen Impfstoffe in ausreichender Menge zur Verfügung. Neu auf den Markt kommende Impfstoffe, die zu den "klassischen" Protein-Impfstoffen zählen, könnten das weiter unterstützen, besonders bei Personen, die den bisherigen Impfstoffen kritisch gegenüberstehen. Besonders bei den zuerst Geimpften, d.h., den Personen, die durch ein fortgeschrittenes Alter und eine Grunderkrankung ein erhöhtes Risiko eines schweren Verlaufs haben, ist der Schutz inzwischen deutlich niedriger, und dadurch wird auch die Belastung der Krankenhäuser durch diese Gruppe wieder zunehmen. Eine rasche Dritt-Impfung dieser recht großen Gruppe, sowie all jener, die Kontakte zu vulnerablen Personen haben, reduziert die erwartete Belastung des Gesundheitssystems deutlich.

Eine deutliche Reduktion der Infektionsdynamik wird aber erst mit Hilfe wirklich flächendeckender Dritt-Impfungen erreicht werden können. Nötig ist hierzu allerdings eine erhebliche Beschleunigung des Impftempos für die Dritt-Impfungen: Da seit Ende Mai ca. 7 % der Bevölkerung pro Woche zweit-geimpft wurden, wäre nun die gleiche Impfgeschwindigkeit bei der Dritt-Impfung 5-6 Monate später möglich und sinnvoll. Simulationen zeigen, dass eine Booster-Kampagne mit dieser Impfgeschwindigkeit bereits nach einem Monat erste Wirkung zeigen wird.

Bis ein ausreichender Teil der Bevölkerung geboostert und geimpft ist, bedarf es in der aktuellen Situation zudem deutlicher und wirksamer Maßnahmen zur Überbrückung, wenn die Intensivstationen nicht stark überlastet werden sollen. Hier sind die bekannten AHA+LA Regeln, sowie konsequent durchgesetzte und flächendeckende Regeln und Testkonzepte im Arbeits- und Freizeitbereich wichtig und hilfreich. Eine vermehrte Testung als alleinige Maßnahme wird zur Durchbrechung der beginnenden Winterwelle wohl nicht reichen. Die Wirksamkeit möglicher Maßnahmen beschreiben wir im Detail im Text. Jedoch sind das nur Überbrückungsmaßnahmen, da sie das Problem der zu hohen Ausbreitungsgeschwindigkeit nicht direkt adressieren, sondern es stattdessen der "natürlichen Immunisierung" durch Infektion überlassen. Die "natürliche Immunisierung" schreitet jedoch deutlich langsamer voran als das Impfen es könnte, und ist auch mit einer erheblichen Krankheitslast verbunden. Insofern erscheint ein schnelles und umfassendes Impfen und Boostern als die wirksamste Methode um die aktuelle Welle bald zu brechen und das Pandemiegeschehen nachhaltig zu kontrollieren.

# Einleitung

Der Winter steht vor der Tür. Im Vergleich zum vergangenen Jahr sind wir in einem wesentlichen Punkt besser aufgestellt: Wir haben sehr wirksame Impfstoffe zur Verfügung, die Mehrzahl der Erwachsenen ist geimpft oder genesen. Trotzdem steigen die Fallzahlen derzeit wieder stark an. Woran liegt das? Was bedeutet das für die kommenden Monate? Und wie können wir damit umgehen? Wir wissen inzwischen wesentlich mehr über das Virus und seine Ausbreitung. Trotzdem gibt es eine Reihe von Missverständnissen und Mythen. Wir möchten deswegen im Folgenden einige der Missverständnisse aus dem Weg räumen und den aktuellen Stand des Wissens zusammenfassen. Auf dieser Basis besprechen wir mögliche Szenarien und diskutieren dann einige ausgewählte Werkzeuge, um gut über den Winter zu kommen.

# Faktengrundlage

## Inzidenz

**Bisher war in Deutschland nur ein kleiner Teil der Bevölkerung mit SARS-CoV-2 infiziert.** Um abzuschätzen, wie sich die Pandemie entwickelt, ist es wichtig zu wissen, wie viele Menschen schon COVID-19 hatten. Rund 5,5 % der Bevölkerung hatten eine bestätigte SARS-CoV-2 Infektion. Berücksichtigt man die unentdeckten Fälle, dann hatten rund 10-15 % der Bevölkerung direkten Kontakt mit dem Virus.[1,2] Das bedeutet gleichzeitig, dass man im eigenen Familien- und Freundeskreis wahrscheinlich nur wenige Personen kennt, die infiziert waren, einen schweren Verlauf hatten oder verstorben sind. Dadurch entsteht manchmal der Eindruck, die Pandemie sei weit weg. Aber genau das ist zu erwarten: Die Wahrscheinlichkeit nach einer Infektion an COVID-19 zu versterben, ist in Deutschland etwa 1 %. Diese 1 % sind auch die Wahrscheinlichkeit für eine 60-jährige Person, nach einer Infektion an COVID-19 zu versterben. Hat eine Person also 100 Freunde und Bekannte, dann sind darunter typischerweise 10 Infizierte, und im Erwartungswert 0,1, also meist keine verstorbene Personen. Viele haben aber von Freunden gehört, dass in deren Familien- oder Freundeskreis jemand verstorben ist. Nimmt man an, dass die "Freunde von Freunden" 10.000 Menschen sind, dann erwartet man in dieser Gruppe 10 Todesfälle. Hätte man die Pandemie im letzten Winter durch deutliche Kontaktbeschränkungen und Vorsichtsmaßnahmen nicht so stark abgeschwächt, hätte es - ähnlich wie in England - deutlich mehr Todesfälle gegeben.

---

[1] Gornyk et al, Deutsches Ärzteblatt 2021, https://www.aerzteblatt.de/pdf.asp?id=221932:
Die Zahl unentdeckter Infektionen war in der ersten Welle etwa 3-6 mal so hoch wie die Zahl der bestätigten Infektionen. In den darauffolgenden Monaten wurde die Teststrategie deutlich ausgeweitet. Seit Sommer 2020 ist der Anteil unentdeckter Infektionen deshalb deutlich geringer: Im Mittel nimmt man aktuell an, dass ⅓ bis ½ der Infektionen unentdeckt bleibt; bei hoher Inzidenz ist diese sogenannte Dunkelziffer etwas höher, bei niedriger auch eher geringer.

[2] https://www.rki.de/DE/Content/Infekt/EpidBull/Archiv/2021/Ausgaben/37_21.pdf?__blob=publicationFile

**Immunisierung der Bevölkerung und "Herdenimmunität".** Zu Beginn der Pandemie sind wir davon ausgegangen, dass der R-Wert unter Eins sinken wird, wenn ungefähr 70% der Bevölkerung immun sind. Das wird manchmal auch als "Herdenimmunitätsschwelle" bezeichnet. Warum ist das jetzt doch nicht der Fall? Der Grund liegt darin, dass die jetzt vorherrschende Delta-Variante sehr viel ansteckender ist als die Ursprungsvariante zu Beginn der Pandemie und dass auch Geimpfte oder Genesene - wenn auch seltener - sich noch infizieren und das Virus weitergeben können. Es müssen also deutlich mehr Personen immun sein (sei es durch Impfung oder nach einer Infektion), um auch ohne Aufrechterhaltung von zusätzlichen Maßnahmen die Inzidenz niedrig zu halten.
Diese Situation hat weitreichende Konsequenzen. Zunächst betrifft das die Ungeimpften. Denn die meisten aus dieser Gruppe werden sich unweigerlich anstecken. Sind weiterhin 20-30% der Bevölkerung nicht geschützt, kommt es in dieser Gruppe zu entsprechend vielen schweren Verläufen und Todesfällen. Zudem bleibt das Gesundheitssystem wegen COVID-19 dauerbelastet (und je nach zeitlichem Verlauf auch überlastet), so dass auch die Versorgung aller anderen Patient*innen eingeschränkt wird. Spätestens bei einer deutlichen Überlastung der Krankenhäuser werden weitergehende Einschränkungen notwendig - mit den bekannten Konsequenzen für unsere Gesellschaft.

**Schutz nach durchgemachter Infektion hat in Deutschland bisher nur geringfügig zur Herdenimmunität beigetragen.** Die heftigen Infektionswellen während der letzten ca. 1,5 Jahre haben die Krankenhäuser erheblich belastet und in dieser Zeit regional zu einer deutlichen Übersterblichkeit geführt - obwohl sich nur 10-15 % der Bevölkerung infiziert haben (siehe oben). Das zeigt, dass eine sogenannte "natürliche Immunisierung" der Bevölkerung - also eine Immunisierung als Folge einer SARS-CoV-2-Infektion - keine realistische Option zur Erreichung einer Herdenimmunität gewesen wäre. Hingegen können wir mit Impfungen 5 % der Bevölkerung *je Woche* immunisieren. Für die Mehrheit derjenigen Menschen, die sich dauerhaft gegen eine Impfung entscheiden, wird es unweigerlich zu einer Infektion kommen, die mit den bekannten altersentsprechenden Risiken für schwere Verläufe, Komplikationen und Tod einhergeht. Eine Überlastung des Gesundheitssystems ist damit in erheblichem Maße davon abhängig, wie viele Personen geimpft sind und in welchem Zeitraum sich die bisher ungeimpften Menschen - und insbesondere Ungeimpfte über 60 Jahre - infizieren.

**Inzidenzen und Neuaufnahmen auf die Intensivstationen sind weiterhin sehr eng korreliert.** Aktuell werden etwa 0,8-0,9 % aller positiv getesteten Personen intensivpflichtig. Dieser Faktor ändert sich nur langsam mit der Zeit und lässt sich gut aus den Daten schätzen. Dadurch ist die Inzidenz ein hilfreicher Frühwarnindikator, selbst wenn man die Dunkelziffer oder Impfquote nicht ganz genau kennen sollte.

# Eingeschränkte Kapazität auf Intensivstationen

**Hohe Inzidenzen werden die Intensivstationen überlasten.** Trotz des Impffortschritts gibt es noch mehr als 3 Millionen Menschen über 60 Jahre und über 11 Millionen zwischen 18 und 59, die nicht geimpft sind. Ein kleiner Anteil hiervon ist genesen. Allerdings lässt die Immunität sowohl bei Geimpften als auch bei Genesenen mit der Zeit nach. Dadurch besteht eine sehr hohe Anzahl Menschen, für die eine SARS-CoV-2-Infektion gefährlich ist. Im Falle einer Infektion muss zum Beispiel eine ungeschützte Person im Alter von 60 mit 2-3 % Wahrscheinlichkeit auf einer Intensivstation behandelt werden. Eine Person, die 20 Jahre jünger ist, hat ein knapp 10 mal geringeres Risiko. Wenn sich *alle* bisher ungeschüzten Menschen in diesem Winter infizieren würden, hätten wir mindestens 3 mal mehr Personen auf den Intensivstationen als es im gesamten letzten Winter der Fall war. Gleichzeitig sind die Reservekapazitäten auf den Intensivstationen dadurch, dass das Personal dauerhaft überlastet ist, die Kliniken verlassen hat oder die Arbeitszeit reduziert hat, in diesem Winter deutlich geringer als im letzten Winter. Durch Personalmangel und Meldekorrekturen stehen rund 4000 Betten weniger zur Verfügung als im letzten Jahr. Wenn die aktuelle Welle also nicht abgebremst wird, dann werden die Intensivstationen für Wochen überlastet und viele andere, notwendige Operationen und Behandlungen müssen verschoben werden.

**Die Intensivstationen haben wenig freie Kapazitäten.** Schon im "Normalbetrieb", also auch ohne COVID-19, sind die Intensivstationen ausgelastet. In Deutschland hat eine typische Intensivstation im Median 12 Betten. Daher ist eine Belegung von mehr als 90 % als kritisch zu betrachten, denn 90 % Belegung heißt, dass typischerweise nur noch *ein* betreibbares Bett frei ist. Diesen Puffer braucht es mindestens*,* damit unvorhersehbare und dringende Fälle wie z.B. Unfallopfer und Menschen mit akutem Herzinfarkt oder Schlaganfall sofort versorgt werden können. Muss ein Krankenhaus aus Kapazitätsgründen einen Notfall ablehnen, kann das Verzögerungen in der Akutversorgung und einen langen Transport bedeuten. Regionale Überlastungen verschärfen das Problem.

**Im Jahr 2020 gab es Engpässe auf den Intensivstationen.** Es wurde berichtet, dass im Jahr 2020 im Jahresmittel nur 3 % der Intensivbetten durch COVID-19 Patient*innen belegt waren. Dieser Durchschnittswert ist jedoch irreführend. Während es in den Monaten Januar, Februar und Mai bis Oktober kaum intensivpflichtige Fälle gab und im März und April nur das südliche Bayern und Baden-Württemberg betroffen waren, stiegen die Belegungszahlen ab November überall dramatisch an. Die Extrembelastung der Monate November und Dezember rechnerisch auf das ganze Jahr 2020 zu verteilen, ist also für die Beurteilung des derzeitigen Geschehens wenig hilfreich. Beim aktuell sehr dynamischen Infektionsgeschehen drohen sehr viele Patient*innen in kurzer Zeit anzufallen.

**Der Pflegemangel bestimmt den Engpass auf den Intensivstationen, nicht ein Bettenmangel.** Ein Bett auf einer Intensivstation ist nur betreibbar, wenn ausgebildetes

Fachpersonal zur Verfügung steht. Der Personalmangel auf den Stationen stellt den eigentlichen Engpass dar und ist nicht schnell und einfach ausgleichbar. Die Intensivpflegeausbildung ist aufwendig und langwierig. Hinzu kommt, dass ein signifikanter Teil des Personals durch die Dauerbelastung ausgebrannt ist. Insofern ist es nicht zielführend, "freie" Betten zu zählen. Es geht in erster Linie darum, wieviel Patient\*innen langfristig ohne Überlastung des Personals betreut werden können.

**Grippe, RSV und andere Erreger bringen zusätzliche Belastung.** Im Winter 2020/21 gab es aufgrund der Maßnahmen gegen die Ausbreitung von SARS-CoV-2 kaum Infektionen mit Influenza oder mit anderen Atemwegserregern, die potenziell schwere Erkrankungen verursachen können. In der Spitze der letzten starken Influenzawelle 2018 lagen zeitgleich etwa 3.000 Patient\*innen auf der Intensivstation. Sollte sich dies im Winter 2021/22 wiederholen, käme dies noch zu der aktuell stark steigenden Zahl an COVID-19 Erkrankten hinzu.

Für Frühgeborene, Säuglinge und Kleinkinder führt eine Infektion mit dem Respiratorischen Synzytial Virus (RSV) und anderer Erreger aus der Gruppe der Paramyxoviren und Picornaviren relativ häufig zu symptomatischen und auch schwereren Verläufen mit notwendiger Sauerstoff- oder sogar Intensivbehandlung. Da RSV- und andere Infektionen der Säuglinge und Kinder im letzten Jahr nicht stattgefunden haben, holen jetzt viele Kinder Erstinfektionen nach.[3] Dies führt bereits aktuell zu einer sehr deutlichen Belastung der pädiatrischen Gesundheitssysteme, während der Effekt von COVID-19 hier geringer ist.

Die AHA+LA Maßnahmen, die gegen die Ausbreitung von SARS-CoV-2 schützen, reduzieren noch effektiver die Ausbreitung von Influenza, RSV und anderer Viren und können diese Wellen auch im kommenden Winter reduzieren oder - wie in 2020/21 weltweit - sogar komplett vermeiden. Eine Erstinfektion mit RSV lässt sich dabei - anders als die Grippeerkrankungen dank Impfung - nicht grundsätzlich verhindern, sondern nur verzögern. Genau da setzen die AHA+LA Maßnahmen an.

## Ausbreitung

**Der doppelte Nutzen der Impfung.** Impfungen gegen COVID-19 schützen nicht nur die geimpfte Person vor Ansteckung und schwerer Erkrankung. Sie reduzieren auch die Übertragung des Virus von Geimpften auf deren Kontaktpersonen, da geimpfte Personen sich seltener anstecken und typischerweise für kürzere Zeit infektiös sind. Es ist für die gesamte Übertragungskette, und für die Verhinderung der Überlastung des Gesundheitssystems wichtig, beide Schutzwirkungen zu beachten.

**Der Schutz gegen Ansteckung ist sehr gut, lässt aber mit der Zeit nach**. Anfangs beträgt der Schutz durch Impfung rund 90 %, das heißt, eine geimpfte Person hat eine circa 10 mal geringere Wahrscheinlichkeit, sich selbst anzustecken. Selbst wenn sich die geimpfte Person ansteckt, ist das Risiko, dass diese das Virus auf eine andere Person überträgt, nochmal zusätzlich reduziert. Insgesamt ergibt sich dadurch für kürzlich Geimpfte eine ca. 20-fach geringere Wahrscheinlichkeit das Virus zu übertragen als für ungeimpfte Personen. Das ist sehr viel. Nach rund 5 Monaten liegt der Schutz durch die Impfung noch bei 50 % bis 70 %, die Wahrscheinlichkeit, sich anzustecken ist also nur noch um etwa einen Faktor 2-3 reduziert. (Die genauen Werte hängen von Impfstoff, Alter, Virusvariante und anderen Aspekten ab.) Das Nachlassen des Impfschutzes ist einer der Gründe, warum die Fallzahlen derzeit steigen. Auch der Schutz gegen einen schweren Verlauf lässt mit zunehmendem Abstand zur Impfung nach, wenn auch auf höherem Niveau (anfangs rund 98 % bei Comirnaty, nach 5 Monaten 90 %, er sinkt also von Schutzfaktor 50 auf einen Schutzfaktor von 10).[4] Dieser Schutz kann durch eine dritte Impfung deutlich erhöht werden. Auch natürliche Infektionen nach einer Impfung können den Schutz verbessern. Genesene Personen haben wahrscheinlich einen ähnlich guten Schutz vor einer Infektion wie Geimpfte. Durch eine Impfung von Personen, die bereits eine SARS-CoV-2-Infektion durchgemacht haben, lässt sich der Schutz nochmal steigern, so dass dieser dann auch den Schutz bei vollständig geimpften Personen deutlich übersteigt.

---

[3] Buda et al, ARE-Wochenbericht zur 43. Kalenderwoche 2021,
https://influenza.rki.de/Wochenberichte/2021_2022/2021-43.pdf

**Ungeimpfte Personen, aber auch diejenigen mit nachlassendem Immunschutz, tragen zur Pandemie bei.** Keine Impfung schützt zu 100 %, auch die gegen COVID-19 nicht. Das ist genauso wie kein Medikament mit 100 % Sicherheit wirkt, oder – an einem Alltagsbeispiel verdeutlicht – der Airbag im Auto nicht 100 % vor einer Verletzung im Fall eines Unfalls schützt. Es ist möglich, dass eine geimpfte Person sich ansteckt, wenn auch weniger wahrscheinlich als bei einer ungeimpften Person. Die Zahl der Durchbruchsinfektionen wird häufig als ein Versagen der Impfung interpretiert. Hierbei wird jedoch nicht beachtet, dass die Gruppe der Geimpften deutlich größer als die Gruppe der Ungeimpften ist. Die Schutzwirkung der Impfung ist weiterhin erheblich. Laut Robert-Koch-Institut beträgt der Schutz vor Krankenhausaufnahme in der Kalenderwoche 40-43 2021 ca. 89 % (Alter 18-59 Jahre) bzw. ca. 85 % (Alter ≥60 Jahre) und der Schutz vor Behandlung auf Intensivstation ca. 94 % (Alter 18-59 Jahre) bzw. ca. 90 % (Alter ≥60 Jahre).[5] Wenn sich eine ungeimpfte Person infiziert, führt das im Mittel zu einer etwa 3-10 mal höheren Belastung der Krankenhäuser, als wenn sich eine Person trotz Impfung infiziert. Man sieht das auch in den nach Impfstatus separat dargestellten aktuellen Inzidenzen. Deshalb sprechen Krankenhäuser auch manchmal von einer Pandemie der Ungeimpften. Diesen Effekt sieht man in allen Altersklassen.
Gleichzeitig würde die jüngere Hälfte der Bevölkerung die Intensivstationen alleine nicht überlasten, da die Wahrscheinlichkeit, schwer zu erkranken, je 20 Jahre an Alter auch um einen Faktor 3-10 zunimmt. Für die Krankenhausbelastung sind 20 Jahre Altersunterschied also ähnlich wie der Unterschied zwischen geimpft und nicht geimpft. Könnte man daraus schliessen, dass es egal ist, ob jüngere Personen sich infizieren?

---

[4] Barda et al, The Lancet 2021, https://www.thelancet.com/journals/lancet/article/PIIS0140-6736(21)02249-2/fulltext und Tartof et al, The Lancet 2021, https://doi.org/10.1016/S0140-6736(21)02183-8
[5] Robert-Koch-Institut, Wöchentlicher Lagebericht vom 04.11.2021,
https://www.rki.de/DE/Content/InfAZ/N/Neuartiges_Coronavirus/Situationsberichte/Wochenbericht/Wochenbericht_2021-11-04.pdf?__blob=publicationFile
Diese Zahlen gelten spezifisch für diese Kalenderwochen.

Abgesehen vom ethischen Aspekt, dass Kinder unter 12 Jahren noch keine gute Möglichkeit haben sich zu schützen, ist ein zweiter Aspekt sehr wichtig: Für die Ausbreitung, und damit für die Inzidenz *in allen Altersgruppen,* spielt die Impfung eine entscheidende Rolle.

**Ein Stabilisieren der Fallzahlen ist bei hoher Inzidenz noch schwieriger als bei niedriger Inzidenz.** Manchen Menschen denken intuitiv, dass wir mehr Kontakte (oder "Freiheiten") haben können, wenn wir in Kauf nehmen, dass die Krankenhäuser voll sind. Das gängige Argument ist, dass man für den Schutz einiger Menschen nicht solche belastenden Einschränkungen akzeptieren sollte. Hier liegt aber ein Missverständnis vor, solange eine Überlastung der Krankenhäuser droht. Lässt man die Fallzahlen steigen, bis die Krankenhäuser überlastet sind, muss man spätestens dann den R-Wert auf 1 stabilisieren. Das Stabilisieren braucht bei hohen Fallzahlen genauso viele Vorsichtsmaßnahmen, wie bei niedrigen Fallzahlen, solange die "natürliche Immunisierung" nur sehr wenig beiträgt. Es geht dann darum, dass der effektive R-Wert 1 sein muss, also dass jede Person nur so viele Kontakte hat, dass maximal eine Person infiziert wird, sollte die Person unwissentlich infiziert sein.

**SARS-CoV-2 wird aller Wahrscheinlichkeit nach endemisch.** Über kurz oder lang wird also im Prinzip jede Person mit dem Virus in Kontakt kommen. Die Entscheidung, die jede Person trifft, ist, ob sie sich vorher durch Impfung schützt oder nicht. Dadurch, dass bisher nicht genug Menschen geschützt sind, kann man die Vorsichtsmaßnahmen noch nicht vollständig aufheben, ohne die Krankenhäuser deutlich zu überlasten.
Muss also in Zukunft jeden Herbst aufs Neue geimpft werden? Das wissen wir noch nicht. Es hängt davon ab, wie lange der Schutz gegen einen schweren Verlauf anhält und ob die Gefahr besteht, dass zu viele Personen gleichzeitig erkranken.

# Szenarien

Zur Pandemiebekämpfung stehen inzwischen viele Instrumente zur Verfügung, welche weiter unten im einzelnen diskutiert werden. Im Folgenden werden daraus einige mögliche Szenarien entwickelt, die exemplarisch mögliche Kombinationen von Instrumenten und daraus resultierende Konsequenzen aufzeigen.

**Szenario "Weiter so!"**
*Annahmen:* Die Inzidenz steigt weiter mit einem R-Wert von 1,2 und bremst sich nur sehr langsam durch weitere freiwillige Einschränkungen und vermehrte Vorsicht einzelner Menschen ab. Es gibt keine anderen Maßnahmen als im Oktober 2021 in Deutschland. Das Impfen und "Boostern" durch Dritt-Impfungen schreitet mit der aktuellen Rate von knapp 1,5 % geimpften Personen pro Woche voran.
*Konsequenzen*: Die Wochenzidenz steigt auf viele hundert je 100.000. Die Impf- bzw. Boosterquote ist nach 7 Wochen um +10 % höher. Die "natürliche Immunisierung" erreicht insgesamt nach sieben Wochen rund +10 % (variabel, abhängig von der Inzidenz). Diese verbesserte Immunität (+20 %) in der Bevölkerung wird durch eine nachlassende Immunantwort in den anderen Teilen der Bevölkerung reduziert. Die Immunität ist also insgesamt etwas verbessert. Ob sie ausreicht, um die Infektionsdynamik deutlich zu reduzieren, wird regional verschieden sein. In Regionen, die schon jetzt 80 % Geimpfte haben, könnte es ausreichen. Trotzdem sind die Krankenhäuser vor allem wegen der vulnerablen und ungeschützten Personen überlastet. Es kommt lokal wahrscheinlich zu Verzögerungen und Engpässen bei der Krankenversorgung und Versorgung der Notfälle. Spätestens dann muss man über eine Änderung der Strategie nachdenken.
*Kurz*: Nicht-Entscheiden ist auch eine "Strategie" und würde wahrscheinlich zu einer Überlastung des Gesundheitssystems führen.

**Szenario "Die Belastungsgrenze des Gesundheitssystems"**
*Annahmen:* Als Maßgabe für die Maßnahmen wird die Belastungsgrenze des Gesundheitssystems genutzt. Das heißt, zusätzlich zu den Hygienemaßnahmen wie AHA+LA, 2G/3G gibt es weitere Einschränkungen, evtl. lokal angepasst, wenn das Gesundheitssystem überlastet ist. Tritt eine Entlastung ein, werden diese Maßnahmen wieder gelockert.
*Konsequenzen*: Die Wocheninzidenz steigt weiter an, bis die zusätzlichen Maßnahmen ihre Wirkung zeigen, kann aber schlussendlich stabilisiert und evtl. auch reduziert werden. Die Inzidenz liegt typischerweise bei einigen hundert Infizierten je 100.000. Es gibt einen gewissen Booster- und Impffortschritt, sowie etwa 1 % natürliche Immunisierung je Woche. Ist das Boostern und Impfen zügig, kann es nach einigen Wochen ausreichen, um die Maßnahmen (regional) etwas zu lockern. Die Krankenhäuser sind deutlich und langfristig belastet und es kann lokal zu Engpässen kommen. Die genaue Entwicklung hängt stark von der Umsetzung der Maßnahmen ab. Wenn die Prämisse ist, erst am Limit der Gesundheitsversorgung deutlich die Maßnahmen zu verstärken, dann ist die Regulierung schwierig, da es keinen Spielraum gibt.
*Kurz*: Insbesondere der Impf- bzw Boosterfortschritt bestimmt die notwendigen Maßnahmen. Bis die Immunisierung ausreichend hoch ist, können verstärkte Maßnahmen und Teststrategien die Inzidenz stabilisieren. Die Belastung im Gesundheitssystem ist jedoch beträchtlich und wird wegen der großen Anzahl ungeschützter Personen, der langen Liegezeiten gerade von jüngeren Patient*innen und wegen der nur langsam nachlassenden Inzidenz auch noch eine Weile hoch bleiben.

**Szenario: "Impf- und Booster-Offensive"**
*Grundlage:* Da die Wirkung der Impfung gegen einen schweren Verlauf und gegen Ansteckung mit der Zeit nachlässt, sind derzeit insbesondere die vulnerablen Personen, die früh geimpft wurden, nicht mehr so gut geschützt. Eine Dritt-Impfung erhöht den Schutz gegen schweren Verlauf und Ansteckung nochmal um ca. einen Faktor 10. Das ist beträchtlich. In Israel haben sich rund 50 % der Menschen boostern lassen. Das war ausreichend, um die Welle dort zu brechen. In Deutschland könnte man sehr wahrscheinlich eine ähnliche Wirkung erzielen. Je schneller man also boostert, desto früher kann die Welle gebrochen werden. Die aktuelle Impfgeschwindigkeit reicht dafür bei weitem nicht aus. Würde es gelingen, rund 7 % der Bürger*innen pro Woche zu boostern, wären vor Weihnachten 50 % der Menschen wesentlich besser geschützt. Solange nicht ausreichend Impfmöglichkeiten bestehen, empfiehlt die STIKO eine klare Priorisierung. Zur Beschleunigung der Impfungen helfen sehr niedrigschwellige Impfangebote, mobile Impfteams, Impfen bei Betriebsärzten, sowie möglicherweise Impfangebote in Geschäften oder Apotheken, begleitet von klarer Kommunikation auch über lokale Influencer. Um 50 % zügig zu erreichen, müsste die sechs-Monatsgrenze der Auffrischung weniger strikt ausgelegt werden. Parallel zum Boostern kann auch die noch nicht geimpfte Bevölkerung niedrigschwellig erreicht werden.

*Wirkung:* Das Boostern von 50 % der Bevölkerung kann einen wesentlichen Beitrag dazu leisten, die Welle aller Voraussicht nach zu brechen. Die Wirkung wird sich sukzessive in der Inzidenz und auch der Aufnahme auf die Intensivstation zeigen. Der Immunschutz ist dann hoch genug, um mit Basismaßnahmen über den Winter zu kommen.

*Kurz:* Boostern ist eine sehr wirksame Maßnahme zum Brechen der aktuellen Welle.

**Anmerkungen**

In den Szenarien zeigt vor allem das Impfen und Boostern eine nachhaltige Wirkung. Andere Maßnahmen dienen der Überbrückung. Hierbei ist zu beachten:

- Die bekannten AHA+LA Maßnahmen erzielen richtig durchgeführt eine gute Wirkung, aber erscheinen alleine derzeit nicht ausreichend.
- 2G und 3G Beschränkungen werden alleine nicht reichen, wenn sie nicht durch weitere Maßnahmen unterstützt werden. Grund sind u.a. die privaten Besuche und Haushaltkontakte, welche nicht unter 2G/3G fallen. In Österreich wird diskutiert, ab einer bestimmten Inzidenz auch private Besuche für Ungeimpfte weitgehend zu unterbinden; in Simulationen zeigt dies sehr gute Wirkung. Ob dies politisch sinnvoll und/oder staatlich durchsetzbar ist, können wir nicht beurteilen.
- Regelmäßiges massives Testen müsste in der Größenordnung von einem Test pro Person pro Woche stattfinden und eine Kontaktnachverfolgung und Quarantäne beinhalten, um die Ausbreitung deutlich zu reduzieren. Dies ist aber logistisch so aufwendig und kostenintensiv, dass eine Umsetzung nicht realistisch erscheint. Ein gezielter Einsatz im Verdachtsfall, vor Treffen mit vulnerablen Personen, in großen Gruppen oder in Gruppen mit hoher Übertragung (Schüler*innen ohne Impfschutz) kann diese Personen schützen.
- Falls es notwendig sein sollte, eine akute Überlastung des Gesundheitssystems zu reduzieren, kann ein "Not-Schutzschalter" notwendig und hilfreich sein (siehe unten). Es ist wichtig, ihn frühzeitig zu planen und so stark wie möglich durchzuführen, damit sich der Aufwand überproportional auszahlt. Im Prinzip kann er auch angewendet werden, um präventiv die Last im Gesundheitssystem zu reduzieren. Ein halbherziger Not-Schutzschalter verfehlt seine Wirkung.

# Pandemie-Instrumentarium
# für den kommenden Winter

Das wirkungsvollste Mittel gegen das Virus, das wir haben, ist die Impfung. Je mehr Menschen sich impfen bzw. Booster-impfen lassen, umso weniger wird sich das Virus verbreiten und desto weniger werden die Intensivstationen belastet sein. Damit würden umso weniger andere Maßnahmen und Einschränkungen nötig sein. Ist die Impflücke jedoch groß, sind Maßnahmen unumgänglich, um das Gesundheitswesen nicht zu überlasten. Wir erläutern daher im Folgenden eine Übersicht an Maßnahmen und machen deren Mechanismen und Wirksamkeit verständlich. Die meisten Maßnahmen sind bekannt, und wir erheben keinen Anspruch auf Vollständigkeit.

Wir betonen, dass diese Mechanismen und ihre Wirksamkeit stark davon abhängen, wie gut sie umgesetzt werden (können). Ob eine bestimmte verpflichtende Maßnahme gerechtfertigt ist, muss letztendlich in einer demokratischen Güterabwägung bestimmt werden. Allen Maßnahmen, die Einschränkung oder Verbot darstellen, sind gleichzeitig ein Schutz anderer Personen und geben diesen Personen die Freiheit, sich selbst und das Gesundheitssystem zu schützen.

## Impfen und Boostern

**Impfen und Boostern ist ein sehr mächtiges Werkzeug zur Vermeidung schwerer Verläufe.** Für die Boosterimpfung sind drei Gruppen besonders wichtig:

1. Personen mit Immunschwächen. Bei diesen Personen hat die Impfung mitunter keine Immunität ausgelöst, so dass durch eine zeitnahe 3. Impfung überhaupt erstmal ein Schutz aufgebaut werden muss.
2. Ältere Menschen, oder Menschen mit bestimmten Risikofaktoren. Nach Erkenntnissen aus Israel[6] ist für das Risiko an einer Durchbruchinfektion schwer zu erkranken das biologische, nicht streng das kalendarische Alter entscheidend. D.h., Menschen mit bestimmten Vorerkrankungen können in einigen Fällen auch trotz einer zweifachen Impfung schwer an COVID-19 erkranken. Da eben diese vulnerable Gruppe auch als Erste in der Impfkampagne im Frühjahr 2021 geimpft wurde, ist eine Boosterimpfung für sie in der aktuellen Situation besonders hilfreich.
3. Beschäftigte im Gesundheitswesen und in Pflegeberufen. Boosterimpfungen in dieser Gruppe können die Verbreitung des Virus und damit die Übertragung auf die, für die eine Fürsorgepflicht besteht, deutlich reduzieren und sind somit für den Schutz der vulnerablen Personen wichtig.

Eine große Studie aus Israel[7] hat Personen mit und ohne dritte Impfung verglichen, und gezeigt, dass sich die Anzahl Hospitalisierungen nach der dritten Impfung nochmal um einen Faktor von gut 10 reduziert (92 % Wirkung) - im Prinzip in jeder Gruppe. Die gesamte Anzahl an Hospitalisierungen, die vermieden wird, ist insbesondere bei älteren Menschen und bei Menschen mit mehreren Risikofaktoren[8] hoch[9]: Personen mit Risikofaktoren haben ein deutlich über zehnmal größeres Risiko, im Krankenhaus behandelt werden zu müssen, verglichen mit denen ohne; mit der Anzahl an Risikofaktoren steigt dieses Risiko deutlich an. Das zeigt deutlich, dass ein Boostern des älteren Teils der Bevölkerung sowie der Menschen mit Risikofaktoren einen sehr starken Beitrag zur Entlastung der Krankenhäuser liefern kann. In Deutschland könnte der Effekt allerdings etwas geringer ausfallen als in Israel, da in Deutschland der Abstand zwischen erster und zweiter Impfung länger ist. Trotzdem ist von einem sehr starken Effekt auszugehen.

---

[6] Barda et al, The Lancet 2021, https://www.thelancet.com/journals/lancet/article/PIIS0140-6736(21)02249-2/fulltext
[7] Barda et al, The Lancet 2021, https://www.thelancet.com/journals/lancet/article/PIIS0140-6736(21)02249-2/fulltext

**Boostern ist ein sehr mächtiges Werkzeug zur Eindämmung der Pandemie.** Die Boosterimpfung stellt nicht nur den Schutz, der kurz nach der zweiten Impfung vorhanden war, wieder her, sondern übertrifft ihn deutlich. Daher sind Personen nach der Boosterimpfung besser und wahrscheinlich auch länger geschützt. Auch ein Schutz vor Ansteckung und Weitergabe des Virus ist in der Gruppe mit der Boosterimpfung um einen Faktor 10 besser als in der Gruppe der zweifach Geimpften. Israel hat mit einem Booster von 50 % der Bevölkerung die dortige Welle im August sehr zügig und deutlich zurückgedrängt. Eine konsequente Boosterkampagne kann also auch in Deutschland sehr wahrscheinlich eine Welle abbremsen, wenn ausreichend viele Menschen zügig erreicht werden. Allerdings haben wir im letzten Sommer maximal 1 % der Bevölkerung pro Tag geimpft (Okt/Nov 2021: 0,2 % pro Tag). Es würde also mindestens 7 Wochen dauern, bis 50 % erreicht worden sind. Bis die Krankenhäuser entlastet sind, dauert es entsprechend länger. Um die Wirkung der Booster zu maximieren, ist es sehr wichtig, zuerst die oben genannten vulnerablen Gruppen aufzufrischen und insgesamt wieder ein hohes Impftempo zu erreichen.

**Globale Solidarität.** Trotz des hohen Nutzens von Boosterimpfungen bleibt es vor allem wichtig, Ländern, die noch keine hohe Impfrate haben, guten Zugang zu Impfdosen zu gewähren. Für den solidarischen Erwerb weiterer Impfdosen steht das COVAX-Programm der Weltgesundheitsorganisation zur Verfügung. Aber auch darüber hinaus gibt es viel zu tun, um eine gerechtere globale Verteilung von Impfstoffen und anderen Ressourcen zu erwirken.[10]

**Eine höhere Impfquote kann durch niedrigschwellige Impfangebote und gute Kommunikation erreicht werden.** Neben der Booster-Impfung sollte nicht vergessen werden, dass eine Steigerung der Impfquote bei den noch nicht Geimpften von großer

---

[8] Dazu gehören Vorerkrankungen wie z.B. Asthma und Krebs, aber auch Faktoren wie z.B. Schwangerschaft und Übergewichtigkeit.
[9] Das Risiko einer Hospitalisierung war im Studienzeitraum für die drei Altersgruppen
16–39, 30-69 und 70+ jeweils, 7, 105 und 574 (je 100.000) ohne eine Drittimpfung. Mit Drittimpfung war das Risiko gut einen Faktor 10 geringer.
Bei Personen mit unterschiedlicher Anzahl Risikofaktoren ist der Effekt ähnlich gross: Mit null oder 1-2 oder 3+
Risikofaktoren war das Risiko 3 bzw. 82 bzw. 504.
[10] Kienzler und Prainsack, Stiftung Entwicklung und Frieden 2021,
https://www.sef-bonn.org/fileadmin/SEF-Dateiliste/04_Publikationen/GG-Spotlight/2021/ggs_2021-02_de.pdf

Bedeutung ist - insbesondere bei der *älteren Hälfte* der Bevölkerung. Es wird deutlich, dass der weitere Zuwachs bei der Impfquote nunmehr sehr langsam ist. Bei der gegenwärtigen Impfrate ist fraglich, ob diese überhaupt den mit dem Abstand zur Impfung nachlassenden Schutz vor Infektion kompensieren kann. Zu der aktuell niedrigen Impfrate haben sowohl eine unterschiedliche Impfbereitschaft in der Bevölkerung als auch Aspekte der Implementierung der Impfkampagnen beigetragen. Ein Kernaspekt ist es, einen einfachen und niedrigschwelligen Zugang zu Impfungen auch langfristig zu sichern, samt einer überzeugenden Informationsstrategie, die der Diversität der Bevölkerung gerecht wird. Vermehrte Gespräche mit der Hausärztin können offene Fragen klären. Mobile Impfteams, mehrsprachige Informationen in diversen Informationskanälen und Zusammenarbeit mit lokalen Institutionen und Influencern, wie religiöse Gemeinden, Sportvereinen und Hausarztpraxen, sind ein wesentlicher Bestandteil dieser Strategie. Eine klare Kommunikation ist essentiell und ein Vertrauen in die Regierung und Wissenschaft hilfreich.

**Langfristig ist eine *generelle* Impfpflicht nicht notwendigerweise zielführend.** Angesichts der Sorge um überfüllte Krankenhäuser ist es verständlich, dass Rufe nach einer Impfpflicht lauter werden. Damit eine generelle Impfpflicht ihre Wirkung entfaltet, darf man kaum Ausnahmen erlauben, müssen empfindliche Strafen verhängt werden, und die Durchsetzung muss flächendeckend funktionieren. Ist dies nicht der Fall, dann läuft man in der aktuellen polarisierten Situation Gefahr, dass eine generelle Impfpflicht zögerliche Menschen eher abschreckt, und überzeugte Impfgegner auch nicht erreicht - weil diese dann lieber die Strafe bezahlen als sich impfen zu lassen. Es ist also unklar, welche Wirkung eine generelle Impfpflicht hätte. Davon differenziert betrachten muss man eine Impfpflicht für diejenigen, die sich um vulnerable Gruppen kümmern.

**Impfgebot am Arbeitsplatz.** Was es bereits heute in bestimmten Institutionen gibt, ist die Regel, dass nur geimpfte bzw. genesene Personen dort arbeiten dürfen. Dies ist etwa bei Berufen der Fall, die enge körperliche Nähe zu anderen Menschen benötigen. Das gilt insbesondere wenn es sich dabei um vulnerable Gruppen handelt, also etwa im Gesundheits- und Pflegebereich. Diese Erfordernis ist nicht mit einer Impfpflicht gleichzusetzen. Auch wenn Arbeitgebende Mitarbeiter*innen, die sich nicht impfen lassen wollen (und nicht anderweitig eingesetzt werden können), kündigen dürfen, darf man ihnen nicht vorschreiben, sich impfen zu lassen. Das ist ein kleiner, aber wichtiger Unterschied. Dies bringt erhebliche Herausforderungen für den Zusammenhalt am Arbeitsplatz und den Zusammenhalt der Gesellschaft mit sich.

## Tests

**PCR-Tests.** PCR-Tests sind nach wie vor die empfindlichsten und spezifischsten Tests auf SARS-CoV-2. Ein PCR-Test sollte z.B. durchgeführt werden, wenn die Patient*in symptomatisch ist, unabhängig von vorausgegangener Impfung oder Infektion, oder wenn die Patient*in Kontakt zu einem bestätigten COVID-19-Fall hatte. Wegen der hohen Sensitivität, kann PCR-Testung als Präventivmaßnahme vor Treffen Infektionen identifizieren. Wie wirksam die Anwendung von Tests in der Prävention ist, hängt von der Umsetzung ab. Für PCR-Tests, die nicht älter als 24 Stunden sind, kann man davon ausgehen, dass in der Praxis grob 90 % der ansteckenden Personen entdeckt werden (je nach Zeitraum seit Probenahme). Antigen-Schnelltests sind etwas weniger sensitiv. Sie entdecken rund 50-75 % der Personen, die durch einen PCR-Test entdeckt werden, und darunter insbesondere diejenigen mit hoher Viruslast. Insbesondere selbst durchgeführte Schnelltests haben den Vorteil, wenig zu kosten und einfacher verfügbar zu sein.

**Antigen-Schnelltests sind nützlich, bieten aber keine vollständige Sicherheit.** Antigen-Schnelltests können nicht alle Infektionen sicher erkennen. Insbesondere zu Beginn einer Infektion, bei fehlerhafter Durchführung oder asymptomatischer Infektion steigt das Risiko, dass eine Person ein negatives Testergebnis bekommt, obwohl tatsächlich eine Infektion vorliegt (falsch-negativ). Die Nutzenden sollte sich daher nie in falscher Sicherheit wiegen. Dennoch kann man mit Schnelltests bei korrekter Anwendung im Durchschnitt etwa die Hälfte der infizierten Personen entdecken,[11] die auch ein PCR Test entdecken würde. Bei stark infektiösen und frisch symptomatischen Personen ist die Entdeckungsrate durch Schnelltests wahrscheinlich sogar höher. Schnelltests eignen sich also als niedrigschwellige Sicherheitsmaßnahme, um Infektionen zu erkennen, v.a. in Verdachtsfällen.

Würde theoretisch in der *jetzigen* Situation *jede* Person etwa einen Schnelltest pro Woche machen, dann könnte die Pandemie eingedämmt werden. Das zeigt auch das Beispiel der Slowakei, wo Massentests eine Prävalenzreduktion von 80 % erreicht haben.[12] Zudem wäre es natürlich notwendig, dass ein möglicherweise positives Ergebnis zeitnah durch PCR-Tests überprüft wird und die Kontaktnachverfolgung, Isolation und Quarantäne schnell und zuverlässig durchgeführt wird. Die Kontakt-Nachverfolgung ist allerdings bei hohen Inzidenzen vom öffentlichen Gesundheitsdienst nicht zu schaffen. Und ein falsch negatives Ergebnis birgt die Gefahr, dass notwendige Hygiene-Maßnahmen wie das Tragen einer Maske vernachlässigt werden. Auch scheint etwa die Hälfte der Bevölkerung selten oder nie einen Test zu machen. Einer der Gründe sind sozioökonomische Konsequenzen: Wer starke Nachteile am Arbeitsplatz oder im Privatleben befürchtet, hat einen Anreiz, sich nicht zu testen. Um die Testbereitschaft zu erhöhen, müssen sozioökonomischen Risiken zum Beispiel durch konsequenten Arbeitnehmer*innenschutz oder finanziellen Ausgleich verringert werden.

**Kosten und Nutzen einer Teststrategie.** Das Kosten-Nutzen-Verhältnis von Tests hängt von den Kosten eines Tests und der Anzahl entdeckter Fälle ab, die je eingesetztem Test entdeckt werden. Bei einer Wocheninzidenz von rund 100 je 100.000 erwartet man, dass maximal einer von 1.000 Tests echt-positiv ist. Diese Werte decken sich mit der Erfahrung aus der Praxis.[13] Setzt man die Kosten pro Test an, lassen sich die Kosten für jeden entdecken positiven Fall berechnen. Der Vorteil von Tests an Schulen und am Arbeitsplatz ist, dass man die Breite der Bevölkerung erreicht. Von den *positiven* Schnelltests in Schulen wurden gut 60 % durch einen PCR-Test bestätigt.[14] Insgesamt weist das darauf hin, dass Teststrategien in Schulen und am Arbeitsplatz für Übergangsphasen hilfreich sein können, insbesondere bei hoher Inzidenz. Hier helfen auch gut durchgeführte Selbsttests dank ihrer guten Verfügbarkeit.

---

[11] Osterman et al, Medical Microbiology and Immunology 2021, https://www.springermedizin.de/covid-19/diagnostik-in-der-infektiologie/evaluation-of-two-rapid-antigen-tests-to-detect-sars-cov-2-in-a-/18773354

[12] Pavelka et al, Science 2021, https://www.science.org/doi/full/10.1126/science.abf9648

[13] Frankfurter Allgemeine Zeitung vom 02.11.2021, https://zeitung.faz.net/faz/rhein-main/2021-11-02/wenige-positive-tests-an-schulen/683549.html?GEPC=s3

## Die Wirkung von nicht-pharmazeutischen Maßnahmen

**2G- und 3G-Veranstaltungen**: Das Ziel von 2G und 3G ist, die Ansteckungsgefahr bei Treffen zu verringern. Hier sind drei Wirkungen zu unterscheiden.

1. Beide Konzepte, 2G und 3G, haben den Vorteil, dass weniger Infizierte zu einer Veranstaltung kommen. Die Inzidenz auf der Veranstaltung wird dadurch verringert. Bei Geimpften ist die Inzidenz[15] derzeit (Okt./Nov. 2021) etwa 4 mal niedriger als bei Ungeimpften; dadurch reduziert sich die Inzidenz auf einer 2G bzw. 3G Veranstaltung (je nach Impfquote). Ein Schnelltest kann ½ bis ¾ der infektiösen Personen entdecken. Es gibt also bei guter Durchführung der Tests (Test findet ¾ der infektiösen Personen) keinen großen Unterschied zwischen 2G und 3G bzgl. der infektiösen Personen.
2. 2G hat den Vorteil gegenüber 3G, dass im Falle der Teilnahme einer infektösen Person weniger Menschen infiziert werden, da geimpfte oder genesene Teilnehmer eine 3-10 mal geringere Wahrscheinlichkeit haben, sich anzustecken. Der Effekt von 3G im Vergleich zu 2G ist aber nicht der Faktor 3-10, sondern hängt vom Anteil ungeschützter Personen bei der Veranstaltung ab. Nimmt man an, dass ⅓ der Personen ungeschützt sind, dann reduziert sich die erwartete Anzahl Infektionen bei 2G gegenüber 3G etwa um einen Faktor zwei. Allerdings stecken sich auch hier vor allem ungeschützte Personen an, die dann auch noch einen geringeren Schutz gegen einen schweren Verlauf haben.[16]
3. Weniger schwere Verläufe: Unter den *infizierten* Personen hat eine geimpfte Person eine deutlich geringere Wahrscheinlichkeit, einen schweren COVID-19 Verlauf zu haben als eine ungeschützte. Damit reduziert 2G die Anzahl Personen, die einen schweren Verlauf haben, nochmal weiter. Die genaue Effektgröße ist nicht leicht abzuschätzen. Das liegt daran, dass die Gruppen der Geimpften und der nicht-Geimpften sich natürlich auch außerhalb der Veranstaltungen treffen - und dort das Virus auf die ungeschützten Personen übertragen werden kann.

---

[14] Frankfurter Allgemeine Zeitung vom 02.11.2021,
https://zeitung.faz.net/faz/rhein-main/2021-11-02/wenige-positive-tests-an-schulen/683549.html?GEPC=s3
[15] Explizit: Die Wahrscheinlichkeit, dass eine geimpfte Person positiv getestet ist.
[16] Annahme: ⅔ der Personen sing genesen oder geimpft. Für die Rechnung gehen wir davon aus, dass die Inzidenz bei genesenen/geimpften versus ungeschützten+Test dieselbe ist.
Dann ist mit der folgender Reduktion bei der erwarteten Anzahl Infektionen zu rechnen für 2G versus 3G:
Annahme Faktor 3 Reduktion: ⅔ * ⅓ + ⅓ =5/9 = 0.55 → etwas halb so viele Infektionen bei 2G statt 3G
Annahme Faktor 4 Reduktion: ⅔ * 1/4 + ⅓ =3/6 = 0.5 → halb so viele Infektionen bei 2G versus 3G.
Annahme Faktor 10 Reduktion: ⅔*1/10 + ⅓ =12/30 = 0.4 → etwas halb so viele Infektionen bei 2G statt 3G

Drei Aspekte gilt es ferner zu bedenken: (1) Gleich ob 2G oder 3G, damit die Maßnahmen wirken, müssen sie auch konsequent umgesetzt werden. (2) Sind Personen, die weder geimpft noch genesen sind, von den 2G Veranstaltungen ausgeschlossen, dann treffen sich diese ungeschützten Personen möglicherweise in einem anderen Kontext und haben dort ein erhöhtes Infektionsrisiko. (3) Ferner spielt es eine Rolle, in wie vielen Bereichen des Lebens 2G, 3G oder eben keine Einschränkungen gelten, um die Gesamtwirkung abzuschätzen. 3G am Arbeitsplatz hätte den Vorteil, dass die ungeschützten Personen aus der Breite der Bevölkerung regelmässig getestet werden.

Verzichtet man bei einer 2G Veranstaltung auf Masken *oder* sind Veranstaltungen größer, kann sich der Vorteil von 2G gegenüber 3G in Bezug auf Übertragung von Infektionen wieder aufheben.

**2G oder 3G mit Testpflicht für *alle* Teilnehmenden.** Zusätzlich zu 2G oder 3G kann im Prinzip auch von den geimpften und genesenen Teilnehmenden ein Test gefordert werden. Dadurch reduziert sich die Wahrscheinlichkeit eines Ausbruchs im Vergleich zu 2G oder 3G alleine nochmal weiter, da noch weniger Personen infektiös eine Veranstaltung besuchen. Bei 2G ist der Reduktionsfaktor genauso wie die Wirksamkeit des gewählten Tests für geimpfte Personen (siehe Abschnitt zu Tests). Bei der Testung aller (3G+Test) ist ein effektiver Wirkfaktor im Vergleich zu 3G etwas geringer, da ja nur die geimpften und genesenen Personen zusätzlich getestet werden, und hier die Wahrscheinlichkeit eines positiven Testergebnisses niedriger ist. Allerdings müssen die Kosten und der Aufwand der Testungen beachtet werden sowie der Effekt auf die Impfbereitschaft.

**Die Veranstaltungsgröße ist wichtig.** Bei Veranstaltungen mit wenig Durchmischung (wie zum Beispiel feste Sitzplätze oder sehr gute Lüftung) steigt die Anzahl der Infizierten linear mit der Veranstaltungsgröße. Bei starker Durchmischung steigt die Anzahl der Infizierten im Mittel *quadratisch* mit der Anzahl Gäste. Starke Durchmischung heißt, dass alle Personen untereinander Kontakt haben oder sich über Aerosole und schlechte Belüftung eine Ansteckung gleichmäßig auf alle verbreiten kann. Eine Verdopplung der Veranstaltungsgröße kann also zu einer Vervierfachung der zu erwarteten Infektionen führen.[17]

**Innenräume mit hoher Personendichte.** Hohe Ansteckungswahrscheinlichkeiten gibt es vor allem in Innenräumen mit hohen Personendichten. Und dies betrifft nicht nur z.B. Innengastronomie oder Diskotheken, sondern auch das Empfangen von Gästen in privaten Wohnungen, genauso wie Besprechungen und Treffen bei der Arbeit. Es betrifft weniger große Hallen, in denen sich die Personen über die vorhandene Flächen verteilen und die Durchlüftung gut ist. Es ist wichtig, dass solche potentiellen Infektionskontexte soweit wie möglich entschärft werden. Die Maßnahmen dafür sind gut bekannt: Personendichten reduzieren, gut Lüften, vorher Testen und Masken tragen.

---

[17] Das liegt daran, dass sich zwei Wahrscheinlichkeiten bzw Raten verdoppeln, (1) die Wahrscheinlichkeit, dass eine infektiöse Person die Veranstaltung besucht ist verdoppelt, *und* (2) eine Verdopplung der Personen, die sich auf der Veranstaltung infizieren (könnten). Das gilt bei starker Durchmischung, wenn also alle untereinander direkten Kontakt haben und/oder die Aerosole sich so im gesamten Raum verbreiten, dass alle Personen erreicht werden.

In Simulationen[18] finden wir, dass ein Selbsttest vor 40% *aller* Treffen (natürlich gefolgt von Selbstisolation bei positivem Test) die derzeitige Infektionsdynamik zumindest stabilisieren würde, *wenn dabei alle, auch die Geimpften, mitmachen würden.* Z.B. könnten alle Arbeitgebenden Selbsttests in genügender Anzahl wöchentlich aktiv verteilen; manche Arbeitgeber*innen machen dies bereits aus eigener Initiative. Sich bei hohen Personendichten in Innenräumen nur auf die Impfung (ohne Boostern) zu verlassen, reicht hingegen für diesen Winter leider nicht aus.

**Schulen, Kinderbetreuung und Bildungseinrichtungen**. Alle Präventions- und Eindämmungsmaßnahmen tragen dazu bei, Kinderbetreuungs- und Schulschließungen zu vermeiden. Um auch bei offenen Schulen die Ausbreitung des Virus angemessen zu reduzieren, sind eine gute Lüftung, der gezielte Einsatz von Luftreinigern, eine inzidenzabhängige Maskenpflicht in allen Schulformen während des Unterrichtes, AHA Maßnahmen, Tests und Kohortierung wirksam. Bei den Tests sind PCR-Tests in der Praxis sensitiver, sie entdecken etwa doppelt so viele infizierte Kinder. Um flächendeckend PCR-Tests zu ermöglichen, sind Pooltestungen in Form der Lolli-Methode hilfreich. Es lässt sich außerdem zeigen, dass das regelmäßige Testen an Schulen und anschließende Quarantänemaßnahmen bei infizierten Schüler*innen und ihren Familien stark dazu beitragen, die Ausbreitung der Pandemie nicht nur an Schulen sondern in der *gesamten* Bevölkerung zu kontrollieren.

**Not-Schutzschalter**: **Gebündelt sind Maßnahmen ungleich wirkungsvoller, um die Fallzahlen zügig und stark zu reduzieren.**
Falls trotz Impfung eine deutliche Beschränkung von Kontakten notwendig werden sollte, um das Gesundheitssystem zu entlasten, dann ist ein kurzer, intensiver „Not-Schutzschalter" sehr effektiv. Je besser man alle Maßnahmen bündelt, desto kürzer kann er ausfallen und desto weniger Kollateralschäden wird er anrichten. Erreicht man so einen R-Wert von 0,7, dann halbieren sich die Fallzahlen jede Woche. So kann innerhalb von nur zwei Wochen die Inzidenz um einen Faktor 4 reduziert werden. In den kommenden Wochen reduziert sich dadurch die Belastung auf den Intensivstationen. Erreicht der R-Wert jedoch nur 0,9, dann braucht dieselbe Reduktion fast 2 Monate. Sollte also ein solcher Eingriff notwendig werden, dann wäre eine Bündelung von allen Maßnahmen ungleich wirkungsvoller als eine Maßnahme alleine. Dazu gehört (i) Home-Office und engmaschige Testpflicht am Arbeitsplatz, (ii) Reduktion der Gruppengröße in Kindergärten, Schulen und am Arbeitsplatz gleichermaßen, (iii) Schließung/Reduktion von Geschäften, Restaurants, Dienstleistungen und Veranstaltungen, sowie generell (iv) deutliche Reduktion von Kontakten auf der Arbeit, in der Öffentlichkeit und im privaten Bereich.
Für eine maximale Wirksamkeit wären diese Maßnahmen konzertiert und gleichzeitig durchzuführen. Verstärkt werden kann die Wirksamkeit durch Testen (je nach Bereich selbst durchgeführte Tests, Tests am Arbeitsplatz, etc.). Ein solcher konzertierter Eingriff könnte deutlich Zeit gewinnen und die Gesundheitsversorgung vorübergehend entlasten. Ferner erscheint die psychologische und wirtschaftliche Belastung eines solchen kurzen

---

[18] Müller et al, MODUS-COVID Bericht vom 22.10.2021, https://doi.org/10.14279/depositonce-12510

Not-Schutzschalters wesentlich geringer, als die durch leichtere, aber ungleich längere Beschränkungen. Der Lockdown-light im Winter 2020/2021 war im Gegensatz zu einem Not-Schutzschalters weder effektiv noch zielführend.

Wegen der hohen negativen gesundheitlichen und edukativen Folgen für Kinder und Jugendliche sowie der erhöhten Belastungen für Eltern (und hier insbesondere für Mütter) sollten Schulschließungen dabei nur als ultima ratio erwogen werden, es sei denn sie wären für eine Entlastung des *pädiatrischen* Gesundheitssystems notwendig.

Es ist unklar, ob ein Not-Schutzschalter notwendig wird. Aber es wäre trotzdem hilfreich, schon jetzt einen klaren Plan zu entwickeln, ihn frühzeitig anzukündigen und mögliche Kollateralschäden präventiv abzufangen. Wichtig ist es deshalb auch, solche Maßnahmen vorab verfassungsrechtlich und ethisch prüfen zu lassen, und eine klare Kommunikationsstrategie zu haben. Sollte ein Eingreifen notwendig werden, dann sollte man sich nicht auf das Schließen von Schulen und Betreuungseinrichtungen beschränken, da die Belastung stark ist, und da es alleine wesentlich weniger wirksam ist als zusammen mit den oben aufgeführten Maßnahmen. Ein kurzer aber intensiver „Not-Aus" an Kontakten in allen Bereichen für kurze Zeit (zwei Wochen) ist ungleich wirksamer und entlastet das System durch die starke Reduktion deutlich länger als zwei Wochen.

**Masken spielen eine wesentliche Rolle um das Ansteckungsgeschehen zu reduzieren.** Zum einen schützen Masken andere Personen: Der Fremdschutz ist extrem hoch, da die großen Tropfen in der Maske bleiben und die kleinen Aerosole stark reduziert werden. Zudem brechen die Masken den Ausatem-Jet und verteilen die Aerosole diffusiv im Raum. Daher ist es unwahrscheinlich die Ausatemluft einer Person, die eine Maske trägt, direkt einzuatmen. Ganz anders ist es, wenn die Person keine Maske trägt. Bei der Delta-Variante ist das Ansteckungsrisiko im Atem-Jet einer infizierten Person sehr hoch, man kann sich beim Einatmen des Atem-Jets schon nach wenigen Minuten anstecken. Masken führen auch zu einem erheblichen Eigenschutz. Eine Person-zu-Person Übertragung in nächster Nähe, wenn beide eine FFP2 Maske tragen, ist massiv reduziert im Vergleich zu einer Situation ohne Maske (mindestens Faktor 10; bei sehr gutem Sitz der Maske wird eine Ansteckung extrem unwahrscheinlich).

In einem Raum können sich im Raumvolumen infektiöse Aerosole anreichern. Lüften oder Luftfiltern reduziert das Risiko einer Ansteckung um ca. einen Faktor 4. Trägt man noch eine Maske dazu schützt man sich um einen Faktor 10-100. Zur Illustration: Nimmt man an, die Wahrscheinlichkeit sich ohne Lüften und Masken anzustecken ist 100%, dann reduziert sich die Wahrscheinlichkeit mit Lüften auf ca 25%. Trägt man zudem eine Maske, ist sie verschwindend gering.

## Gesundheitssystem

Wie oben erläutert, ist es ein sehr wahrscheinliches Szenario, dass Krankenhäuser und vor allem Intensivstationen diesen Winter erneut an ihre Kapazitätsgrenzen kommen. Um mit einer solchen eingetretenen Situation bestmöglichst umzugehen, sollten bereits im Vorhinein gewisse Vorbereitungen getroffen werden. Denn wenn Ressourcen knapp werden, müssen diese so gerecht und effizient wie möglich eingesetzt werden.

**Flächendeckend umgesetztes Advanced Care Planning entlastet das Gesundheitssystem sowie Patient*innen und ihre Angehörigen.** Mit Advanced Care Planning (ACP) wird die vorausschauende Versorgungsplanung beschrieben. ACP soll es Menschen ermöglichen, sich mit ihrer zukünftigen medizinischen Versorgung auseinanderzusetzen und diese möglichst so zu planen, dass sie den eigenen Erwartungen, Wünschen und Wertvorstellungen entspricht (z.B. über eine Patientenverfügung). Die Vorstellungen zur Versorgung in der letzten Lebensphase sind sehr unterschiedlich; während einige Menschen eine auch invasive Behandlung "bis zum letzten Atemzug" wünschen, gibt es andere, die bestimmte etwa intensivmedizinische Maßnahmen für sich ablehnen.

Aus Patient*innenenperspektive empfiehlt es sich deswegen in jedem Fall, auch während einer Pandemie, klar zu benennen, ob ggf. eine Aufnahme auf eine Intensivstation oder eine invasive Beatmung überhaupt gewünscht wird. Dies erscheint umso dringlicher, je absehbarer wird, dass ggf. Engpässe in der Versorgung entstehen. In einer solchen Situation leidet absehbar die Versorgungsqualität und Entscheidungen sind ggf. nicht mehr nur Patient*innen-zentriert, sondern müssen in der Notlage überindividuelle Entscheidungskriterien berücksichtigen.[19]

# Das Ende der Pandemie

**Müssen wir weiterhin mit noch ansteckenderen Virusvarianten rechnen?**
Die Delta-Variante des SARS-CoV-2 ist eine sehr weit optimierte Variante des Virus, die mehrere Mutationen so vereint, dass sie sehr ansteckend geworden ist. Sie kann dadurch, dass sie sich im Körper schneller ausbreiten kann, auch dem Impfschutz besser entgehen als die ursprüngliche Variante. Es besteht die Frage, ob das Virus sich weiter verändern kann und uns dadurch neue Probleme bereitet.

Ein Virus muss sein Genom immer so stabil halten, dass alle Funktionen, die es zur Vermehrung braucht, aufrecht erhalten werden; d.h., die Möglichkeit an Variationen ist limitiert. Die Tatsache, dass sich immer wieder SARS-CoV-2-Varianten durchsetzen, die ähnliche Veränderungen tragen, zeigt deutlich, dass die Variations-Fähigkeit des Virus limitiert ist, und es nicht möglich ist, endlos neue Varianten auszubilden.

Eine neue Virusvariante wird sich nur durchsetzen, wenn sie einen Vorteil hat, d.h., sich entweder schneller ausbreitet oder der Immunantwort besser entgehen kann. Stärker krank zu machen ist per se kein Vorteil für ein Virus, sondern eher ein Nachteil. Denn diejenigen, die schwerer krank sind, sind weniger mobil und stecken dadurch andere weniger häufig an. Es ist aber sehr gut möglich, dass sich in Zukunft Varianten entwickeln werden, die eine Immunantwort besser umgehen.

---

[19] Deutsche Interdisziplinäre Vereinigung für Intensiv- und Notfallmedizin, S1-Leitlinie vom 17.04.2020, https://www.awmf.org/uploads/tx_szleitlinien/040-013l_S1_Zuteilung-intensivmedizinscher-Ressourcen-COVID-19-Pandemie-Klinisch-ethische_Empfehlungen_2020-07_2-verlaengert2.pdf

**Brauchen wir jedes Jahr eine Auffrischungsimpfung?**
Ob in Zukunft alle Personen oder bestimmte Personengruppen regelmäßig eine Auffrischungsimpfung brauchen, können wir noch nicht absehen. Dafür brauchen wir Daten zu wiederholten Infektionen. Diese Daten liegen derzeit noch nicht vor, aber es ist zu erwarten, dass wir nach diesem Winter hierzu mehr Kenntnisse haben.

**Wann ist die Pandemie vorbei?**
Das ist keine triviale Frage. Man muss hier mindestens drei Dimensionen unterschieden: (1) eine medizinische/virologische/epidemiologische Dimension, (2) eine mentale und habituelle Dimension und (3) eine politisch-rechtliche Dimension. Die medizinische Dimension spricht eine klare Sprache: Die Pandemie ist nicht vorbei. Es scheint so zu sein, dass die pandemische Lage irgendwann verschwindet, die Existenz des Virus aber nicht beendet werden kann und sich COVID-19 zu einer unter anderen Infektionskrankheiten entwickeln wird. Es wird dann weiterhin ein dichtes Monitoring-Netzwerk, eine Reihe von Nachimpfungen und entsprechend Forschung notwendig sein. Das verweist auf die zweite Dimension. Derzeit ist eine nachlassende Aufmerksamkeit im Alltagshandeln dem Virus gegenüber zu beobachten. Lockerungen der Verhaltensstandards und größere Resistenz gegen die Krankheit führt fast automatisch dazu, dass sich sorgloseres Verhalten habitualisiert, was nachvollziehbar und auch wünschenswert ist. Zugleich wird es in dieser Situation schwieriger, Maßnahmen aufrechtzuerhalten oder sogar wieder zu verstärken. Gegen diese Trägheit gesellschaftlicher Routinen muss sich nun das Monitoring der Krise durchsetzen. Der Wunsch, dass die Pandemie weitgehend beendet sei, steht damit der Notwendigkeit gegenüber, die Aufmerksamkeit auf einem hohen Level zu belassen. In der dritten Dimension führt das unweigerlich zu Zielkonflikten und zu inkonsistenter Kommunikation. Die Menschen scheinen es leid zu sein, mit weiteren Maßnahmen konfrontiert zu werden, was zugleich die Notwendigkeit von Maßnahmen erhöht. Dieser Zirkel verweist darauf, dass die Pandemie noch nicht vorbei ist.

Bezogen auf die Zukunft muss erwartet werden, dass die Pandemie zu einem Ende gekommen sein wird, wenn die Maßnahmen gegen die Pandemie (Einführung von lokalen bzw. bereichsspezifischen Schutzmaßnahmen, fortgesetzte Impfungen) so routinisiert sind, dass sie unterhalb der Aufmerksamkeitsschwelle verlaufen. Routinen dienen dazu, Entscheidungen semantisch möglichst wenig sichtbar zu machen.

Das gilt im übrigen auch für die Erhöhung der Impfbereitschaft, bei der einfache Zugänge mindestens eine genauso große Rolle spielen dürften wie Informationskampagnen und professionalisierte Kommunikationsstrategien, wie wir sie u.a. vom Beginn der AIDS-Aufklärung etwa kennen.

Wann eine vollständige Beendigung von nicht-freiwilligen Maßnahmen denkbar ist, ist derzeit nicht abzusehen. Dies in den Winter hinein überhaupt zu diskutieren, ist völlig unrealistisch, muss aber das Ziel für die Zeit nach dem Winter sein. Wie sehr dies von einer erfolgreichen Erhöhung der Impfbereitschaft und der logistischen Planung weiterer Impfwellen abhängig sein wird, sollte deutlich geworden sein.

## Autor*innen

Viola Priesemann, Max-Planck-Institut für Dynamik und Selbstorganisation, Göttingen

Eberhard Bodenschatz, Max-Planck-Institut für Dynamik und Selbstorganisation, Göttingen

Sandra Ciesek, Universitätsklinikum Frankfurt, Goethe-Universität, Frankfurt

Eva Grill, Institut für Medizinische Informationsverarbeitung, Biometrie und Epidemiologie, Ludwig-Maximilians-Universität München (LMU), München

Emil N. Iftekhar, Max-Planck-Institut für Dynamik und Selbstorganisation, Göttingen

Christian Karagiannidis, Lungenklinik Köln-Merheim, Universität Witten/ Herdecke

André Karch, Westfälische Wilhelms-Universität Münster, Münster

Mirjam Kretzschmar, University Medical Center Utrecht, Utrecht, Die Niederlande

Berit Lange, Epidemiologie, Helmholtz-Zentrum für Infektionsforschung, Braunschweig

Sebastian A. Müller, Fachgebiet Verkehrssystemplanung und Verkehrstelematik, Technische Universität (TU) Berlin, Berlin

Kai Nagel, Fachgebiet Verkehrssystemplanung und Verkehrstelematik, Technische Universität (TU) Berlin, Berlin

Armin Nassehi, Institut für Soziologie, Ludwig-Maximilians-Universität München (LMU), München

Mathias W. Pletz, Institut für Infektionsmedizin und Krankenhaushygiene, Universitätsklinikum Jena, Jena

Barbara Prainsack, Institut für Politikwissenschaft, Universität Wien, Wien, Österreich

Ulrike Protzer, Institut für Virologie, Technische Universität München / Helmholtz Zentrum München, München

Leif Erik Sander, Medizinische Klinik mit Schwerpunkt Infektiologie und Pneumologie, Charité - Universitätsmedizin Berlin, Berlin

Andreas Schuppert, RWTH Aachen / Universitätsklinikum Aachen, Aachen

Anita Schöbel, Fraunhofer-Institut für Techno- und Wirtschaftsmathematik (ITWM), Kaiserslautern und Fachbereich Mathematik, TU Kaiserslautern

Klaus Überla, Virologisches Institut, Universitätsklinikum Erlangen, Erlangen

Carsten Watzl, Leibniz Institut für Arbeitsforschung (IfADo), TU Dortmund, Dortmund

Hajo Zeeb, Leibniz Institut für Präventionsforschung und Epidemiologe-BIPS, Bremen